\documentclass{article}

\usepackage[noblocks]{authblk}
\usepackage{amsmath,amssymb,amsfonts}
\usepackage{algorithmic}
\usepackage{cite}
\usepackage{graphicx}
\graphicspath{{img/}}
\usepackage{lscape}
\usepackage{textcomp}
\usepackage{xcolor}

\usepackage{booktabs}
\usepackage[capitalise]{cleveref}
\usepackage{siunitx}
\usepackage{subcaption}
\usepackage{tablefootnote}
\usepackage{url}

\newcommand{\rnum}[1]{\uppercase\expandafter{\romannumeral#1\relax}}
\newcommand{\Rho}{\mathrm{P}}

\begin{document}

\title{Airdrops and Privacy: A Case Study in Cross-Blockchain
  Analysis}

\author{Martin Harrigan\thanks{Email: martin.harrigan@itcarlow.ie}}
\author{Lei Shi\thanks{Email: lei.shi@itcarlow.ie}}
\affil{Institute of Technology, Carlow}

\author{Jacob Illum\thanks{Email: jacob@chainalysis.com}}
\affil{Chainalysis Inc., New York}

\date{}

\maketitle

\begin{abstract}
Airdrops are a popular method of distributing cryptocurrencies and
tokens. While often considered risk-free from the point of view of
recipients, their impact on privacy is easily overlooked. We examine
the Clam airdrop of 2014, a forerunner to many of today's airdrops,
that distributed a new cryptocurrency to every address with a non-dust
balance on the Bitcoin, Litecoin and Dogecoin
blockchains. Specifically, we use address clustering to try to
construct the one-to-many mappings from entities to addresses on the
blockchains, individually and in combination. We show that the sharing
of addresses between the blockchains is a privacy risk. We identify
instances where an entity has disclosed information about their
address ownership on the Bitcoin, Litecoin and Dogecoin blockchains,
exclusively via their activity on the Clam blockchain.

\end{abstract}

\section{Introduction}\label{sec:introduction}

An airdrop is a free distribution of a cryptocurrency or token. It
generally involves a large number of recipients and uses external
systems, for example, existing blockchains or centralised identity
providers, to prevent Sybil attacks. In practice, the parameters of an
airdrop are arbitrarily set by the distributor. Airdrops are often
considered risk-free since the recipients do not part with anything of
value. However, this is not true in the general case. For example, one
method of preventing Sybil attacks uses addresses on existing
blockchains to select the recipients. In order to claim the free
cryptocurrency or tokens, a recipient must reuse his or her existing
private-keys to create a transaction on the blockchain belonging to
the airdrop. The contents of this transaction, in particular the reuse
of addresses across blockchains, its relationship with other
transactions, and the very act of broadcasting the transaction itself
may inadvertently disclose information about the recipients to
interested third parties.

This paper considers the impact of the Clam airdrop on privacy across
three blockchains: Bitcoin, Litecoin and Dogecoin. We use
\textit{address clustering} to try to construct the one-to-many
mappings from entities to addresses on the blockchains, individually
and in combination. An address clustering partitions the set of
addresses observed on a blockchain into maximal subsets of addresses
that are controlled by the same entity. Each component of the
partition is an address cluster. When combined with address tagging,
that is, associating real-world identities with addresses, and graph
analysis, it is an effective means of analysing and deanonymising
blockchain activity at both the micro- and macro-levels, see,
e.g.,~\cite{reid-harrigan-13,meiklejohn-et-al-13,androulaki-et-al-13,ober-et-al-13,spagnuolo-et-al-13,lischke-fabian-16,ermilov-et-al-17}.

Experimental analysis has shown that a single heuristic for address
clustering, the \textit{multi-input heuristic}, can identify more than
\num{69}\% of the addresses in the wallets stored by lightweight
clients~\cite{nick-15}. This heuristic assumes that the addresses
referenced in transaction outputs spent in a single multi-input
transaction are controlled by the same
entity~\cite{nakamoto-08}. Although vulnerable to techniques such as
CoinJoin~\cite{maxwell-13} and its kin, it is a useful heuristic in
practice~\cite{harrigan-fretter-16}. The analyses in this paper use
the multi-input heuristic exclusively but can be extended to any
number of other heuristics, including those that reduce the rate of
false positives~\cite{goldfeder-et-al-17}.

We examine the impact of the address clustering for the Clam
blockchain on the address clusterings for the Bitcoin, Litecoin and
Dogecoin blockchains. In other words, we identify cases where an
entity has disclosed information about their address ownership on the
Bitcoin, Litecoin and Dogecoin blockchains, exclusively via their
activity on the Clam blockchain.

The paper is organised as follows. In \cref{sec:related-work} we
briefly list related work in the areas of cross-blockchain analysis,
and airdrops. In \cref{sec:four-blockchains} we describe the four
blockchains and quantify the sharing of addresses between them. We
analyse the address cluster sizes and coverage and we quantify the
levels of address reuse and address cluster
merging. \Cref{sec:single-blockchain-impact} explores the impact of
the address clustering for the Clam blockchain on its counterparts
individually. \Cref{sec:multi-blockchain-impact} generalises the
approach to handle many blockchains in combination. We conclude with
some future work in \cref{sec:conclusion}.

\section{Related Work}\label{sec:related-work}

We can categorise related work into two areas: cross-blockchain
analysis and airdrops.

In the first category, Nieves~\cite{nieves-18} evaluated heuristics
for recognising cross-blockchain transactions, in particular those
facilitated by centralised services such as ShapeShift and
Changelly\footnote{https://shapeshift.io, https://changelly.com}. The
heuristics involve the identification of pairs of transactions from
different blockchains that can be linked to known service addresses
and have similar timestamps and output values. Popuri and
Gunes~\cite{popuri-gunes-16} performed a network analysis of the
address graphs derived from the Bitcoin and Litecoin blockchains. They
identify power laws in their in- and out-degree distributions, in-out
and out-in degree correlations that vary between negative
(disassortative) and neutral (non-assortative), and global clustering
coefficients that decrease with time. Kalodner et
al.~\cite{kalodner-et-al-17} present BlockSci, an open-source platform
for blockchain analysis that supports many blockchains that are
schematically similar to Bitcoin. In an example usage, the authors
compute a form of velocity for Bitcoin and propose it as a useful
metric for making comparisons across cryptocurrencies. There exist
many comparative studies of the market performance of
cryptocurrencies. For example, ElBahrawy et
al.~\cite{elbahrawy-et-al-17} compare \num{1400} cryptocurrencies by
market capitalisation over a four year period. However, these studies
are only tangentially related to the underlying blockchains.

Although airdrops are a popular method of distributing cryptocurrency,
our second category has few entries. The MIT Bitcoin
Project~\cite{rubin-elitzer-14} organised a pseudo-airdrop by offering
every MIT undergraduate \$\num{100} worth of bitcoin in October
2014. The experimenters used the airdrop to examine the behaviour of
natural early adopters (NEAs)~\cite{catalini-tucker-16}. However, we
are not aware of any studies that examine the mechanisms and the wider
impact of airdrops.

We use common terminology from graph theory through-out the
paper. Please refer to Diestel~\cite{diestel-17} or a similar
reference for definitions.

\section{The Four Blockchains}\label{sec:four-blockchains}

Due to their shared lineage, the Bitcoin, Litecoin, Dogecoin and Clam
blockchains are schematically similar, i.e.\ they have similar block
headers, transaction formats, etc. However, they have very different
sizes. \Cref{tab:four-blockchains} compares the four blockchains based
on their transaction, transaction output, address, address cluster and
\textit{non-trivial} address cluster counts. Non-trivial address
clusters are those that contain more than one address; trivial address
clusters contain exactly one. At the time of this analysis, the Clam
blockchain is small ($\sim 16$~million transaction outputs); the
Litecoin and Dogecoin blockchains are medium-sized ($\sim 84$~million
and $\sim 117$~million transaction outputs respectively); and the
Bitcoin blockchain is large ($\sim 845$~million transaction outputs).

\begin{landscape}
\begin{table*}
  \centering
  \caption{We analyse the Bitcoin, Litecoin, Dogecoin and Clam
    blockchains as of 1st May 2018. The last eight hexadecimal digits
    of the blocks at the tip of each blockchain are shown below. All
    analysis in this paper is based on these
    snapshots.}\label{tab:four-blockchains}
  \begin{tabular}{lrrrr}
    \toprule
    \multicolumn{1}{c}{} &
    \multicolumn{1}{c}{Bitcoin} &
    \multicolumn{1}{c}{Litecoin} &
    \multicolumn{1}{c}{Dogecoin} &
    \multicolumn{1}{c}{Clam}\\
    \midrule
    Tip Hash Ending & \texttt{0x5bd3e36d} & \texttt{0x709b646d} & \texttt{0xe6f50d81} & \texttt{0x31873627}\\
    \# Transactions & \num{313 522 772} & \num{23 844 704} & \num{36 371 905} & \num{5 036 940}\\
    \# Transaction Outputs & \num{845 351 818} & \num{83 998 454} & \num{117 403 235} & \num{15 648 326}\\
    \# Addresses & \num{389 757 330} & \num{28 696 848} & \num{30 573 438} & \num{3 775 072}\\
    \# Address Clusters & \num{183 312 914} & \num{15 876 469} & \num{19 521 831} & \num{3 435 622}\\
    \# Non-Trivial Address Clusters & \num{37 362 066} & \num{1 802 872} & \num{1 355 230} & \num{21 034}\\
    \bottomrule
  \end{tabular}
\end{table*}
\end{landscape}

Clam is the least known of the four cryptocurrencies. Its source code
is a fork of Blackcoin, which is a fork of Novacoin, which is a fork
of Peercoin, which is a fork of Bitcoin. It derives much of its
functionality from its ancestors. Clam was announced on the 24th March
2014.\footnote{https://bitcointalk.org/index.php?topic=623147.0} An
airdrop sent every address with a non-dust balance on the Bitcoin,
Litecoin and Dogecoin blockchains, as of block heights \num{300 377},
\num{565 693} and \num{218 556} respectively, \num{4.6}~CLAM\@.

\begin{figure}
  \begin{subfigure}[c]{\linewidth}
    \centering
    \includegraphics[page=1,width=0.6\linewidth]{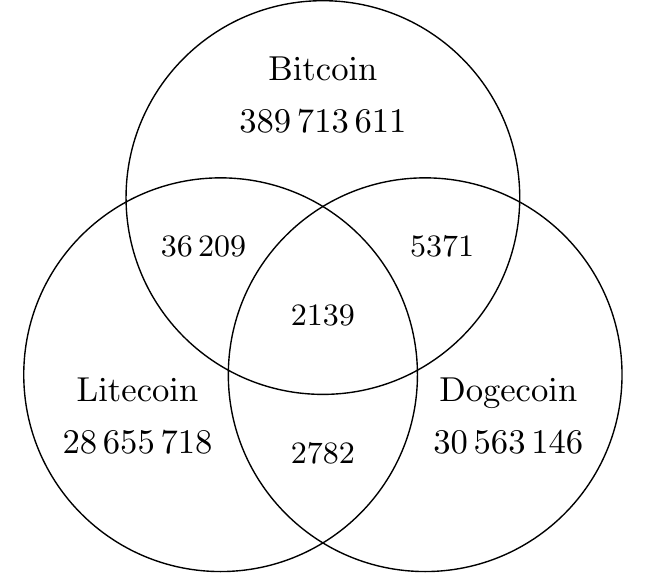}
    \caption{The number of addresses on the Bitcoin, Litecoin and
      Dogecoin blockchains and how they are shared. For example,
      \num{36 209} addresses are shared between the Bitcoin and
      Litecoin blockchains but are not on the Dogecoin
      blockchain. This Venn diagram does not consider the Clam
      blockchain.}\label{fig:addrs-without-clam}
    \vspace{1.0cm}
  \end{subfigure}
  \begin{subfigure}[c]{\linewidth}
    \centering
    \includegraphics[page=2,width=0.6\linewidth]{clam-address-sharing/main.pdf}
    \caption{The number of addresses on the Clam blockchain and how
      they are shared with the Bitcoin, Litecoin, and Dogecoin
      blockchains. In this instance, the universe (the rounded
      rectangle) is the set of addresses on the Clam
      blockchain.}\label{fig:addrs-with-clam}
  \end{subfigure}
  \caption{The sharing of addresses between the Bitcoin, Litecoin,
    Dogecoin and Clam blockchains.}\label{fig:addrs-without-and-with-clam}
\end{figure}
  
The impact of the airdrop on the sharing of addresses between the four
blockchains is illustrated in
\cref{fig:addrs-without-and-with-clam}. We observe that the Bitcoin,
Litecoin and Dogecoin blockchains share a small number of addresses
(see \cref{fig:addrs-without-clam}). However, there is significant
sharing of addresses with the Clam blockchain (see
\cref{fig:addrs-with-clam}). For example, of the \num{3 775 072}
addresses on the Clam blockchain, only \num{566 642} are not shared
with any of the other three blockchains.

\subsection{Address Cluster Size and Coverage}\label{sec:cluster-analysis}

The address cluster sizes and coverage for each blockchain are
visualised in
\cref{fig:bitcoin-cluster-analysis,fig:litecoin-cluster-analysis,fig:dogecoin-cluster-analysis,fig:clam-cluster-analysis}. The
address cluster sizes are binned in the histograms. Both the
horizontal and the vertical axes are log-scaled. The inset pie charts
show the coverage, or total number of addresses in the address
clusters, in each size range. The address clustering for the Clam
blockchain differs from the others: \num{90}\% of it addresses are in
trivial address clusters. This is primarily due to the airdrop; many
of the addresses appear in a single transaction output that was
assigned \num{4.6}~CLAM\@. If we ignore this difference, then we can
observe the significant coverage provided by the large address
clusters: \num{30}\% of the addresses in the Bitcoin blockchain are in
address clusters of size greater than \num{100}; the comparable
numbers for Litecoin and Dogecoin are \num{28}\% and \num{25}\%,
respectively. These are significant when tagging address clusters with
real-world identities: the large address clusters provide good
coverage.

The largest address cluster on the Bitcoin blockchain is an anomaly:
it originally belonged to the Mt.\ Gox exchange that, for a time,
allowed users to import private-keys directly from their wallets. This
feature causes the multi-input heuristic to produce false positives
and requires additional heuristics and information not available on
the blockchain in order to correct for it. For the purposes of this
analysis we will ignore the impact of this false positive.

\subsection{Address \& Address Cluster Novelties}\label{sec:tx-analysis}

\begin{figure*}
  \centering
  \begin{subfigure}[c]{.45\linewidth}
    \includegraphics[width=\linewidth]{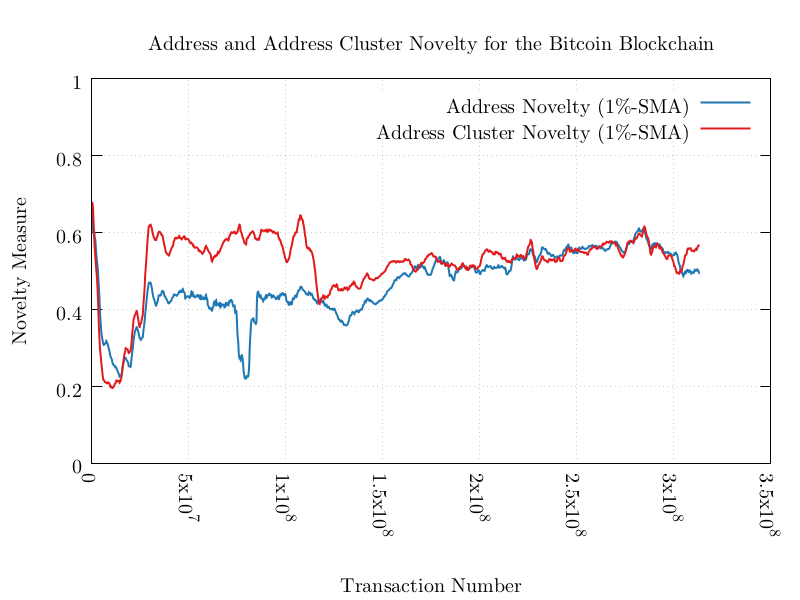}
    \caption{}\label{fig:bitcoin-tx-analysis}
  \end{subfigure}
  \begin{subfigure}[c]{.45\linewidth}
    \includegraphics[width=\linewidth]{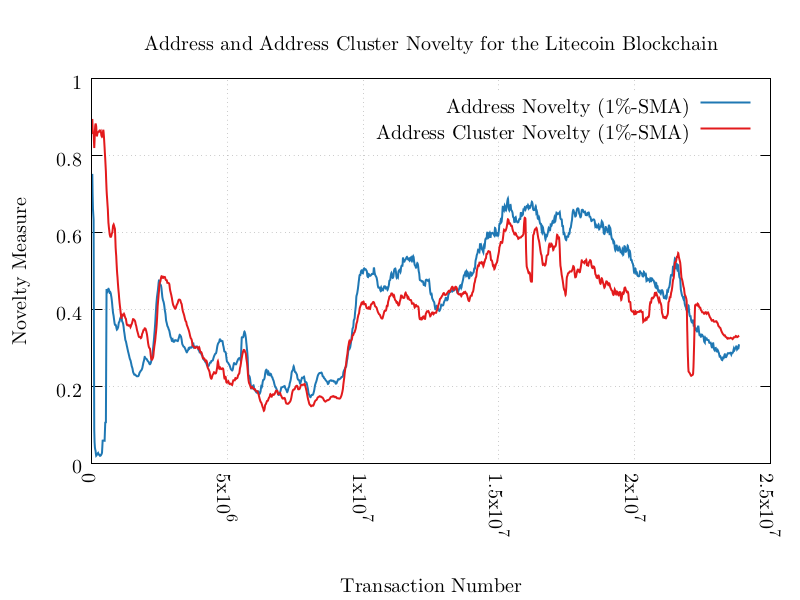}
    \caption{}\label{fig:litecoin-tx-analysis}
  \end{subfigure}
  \begin{subfigure}[c]{.45\linewidth}
    \includegraphics[width=\linewidth]{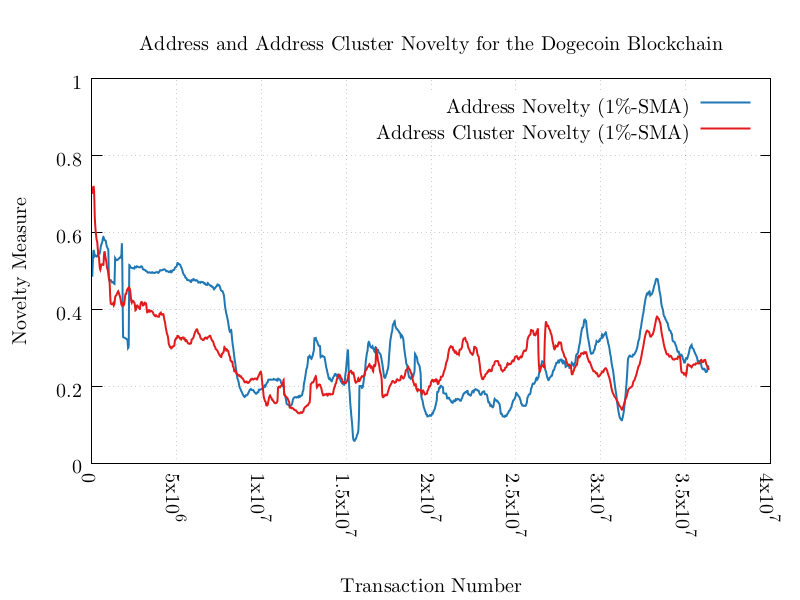}
    \caption{}\label{fig:dogecoin-tx-analysis}
  \end{subfigure}
  \begin{subfigure}[c]{.45\linewidth}
    \includegraphics[width=\linewidth]{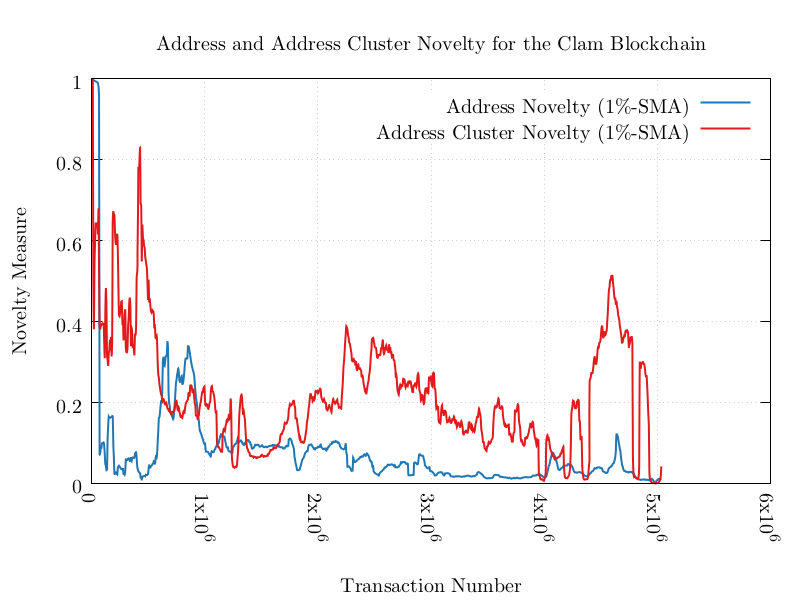}
    \caption{}\label{fig:clam-tx-analysis}
  \end{subfigure}
  \begin{subfigure}[c]{.45\linewidth}
    \includegraphics[width=\linewidth]{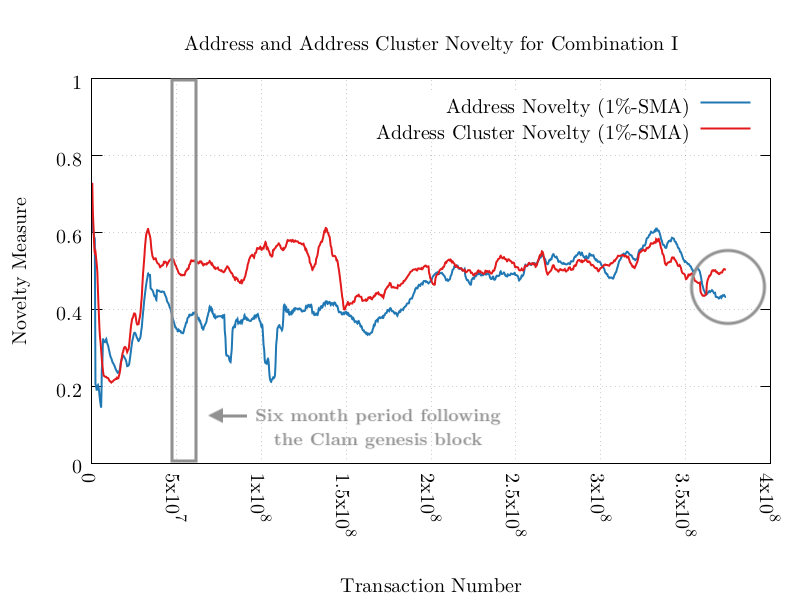}
    \caption{}\label{fig:bitlitedoge-tx-analysis-annotated}
  \end{subfigure}
  \begin{subfigure}[c]{.45\linewidth}
    \includegraphics[width=\linewidth]{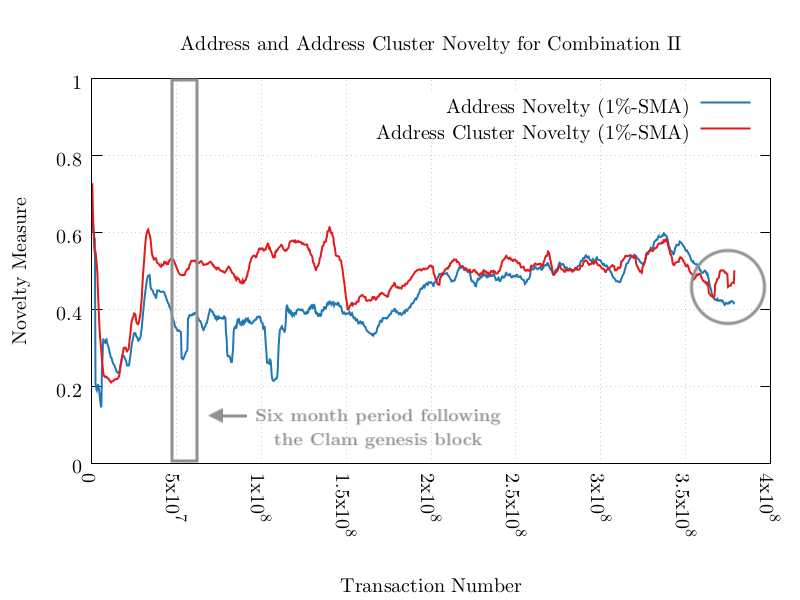}
    \caption{}\label{fig:bitlitedogeclam-tx-analysis-annotated}
  \end{subfigure}
  \caption{Line charts showing the address and address cluster novelties
    for the four blockchains and two combinations.}\label{fig:tx-analysis}
\end{figure*}

The levels of \textit{address novelty} and \textit{address cluster
  novelty} for each blockchain are plotted in
\cref{fig:bitcoin-tx-analysis,fig:litecoin-tx-analysis,fig:dogecoin-tx-analysis,fig:clam-tx-analysis}. We
define the address novelty of a transaction to be its number of
\textit{new addresses} divided by its number of transaction outputs. A
new address is one that has not yet been observed in that
transaction's blockchain. Nakamoto~\cite{nakamoto-08} advised that ``a
new key pair should be used for each transaction.'' This is from the
perspective of the payees only; if the payer requires additional
transaction outputs, say, for change, they should also generate a new
key pair. For transaction outputs that contain \texttt{P2PK} and
\texttt{P2PKH} scripts, the number of transaction outputs in a
transaction is the potential number of new addresses. It can be
adjusted for \texttt{OP\_RETURN} scripts, multi-signature scripts and
\texttt{P2SH} scripts where the redemption script is known. If
everyone followed Nakamoto's advice, the address novelty for every
transaction would be $1$. However, due to address reuse, this is not
the case.

We plot the address novelties for each blockchain using a simple
moving average (SMA) that includes the last \num{1}\% of transactions
(blue lines). To adjust for the variable rate of transactions across
the blockchains, we place the ordinal transaction number, rather than
time, along the horizontal axis. This compresses periods of
low-activity and expands periods of high-activity. We observe that the
address novelty in the Dogecoin blockchain
(\cref{fig:dogecoin-tx-analysis}) is much lower than that in the
Bitcoin blockchain (\cref{fig:bitcoin-tx-analysis}). The Clam
blockchain has a high address novelty during the airdrop but low
values elsewhere (\cref{fig:clam-tx-analysis}). If addresses are
shared between blockchains then the low address novelties for the
Dogecoin and Clam blockchains may be weak links.

Similarly, we define the address cluster novelty of a transaction
whose transaction inputs reference at least two different addresses to
be a Boolean value that is one if the transaction merges two or more
trivial address clusters, and zero otherwise. The metric is only
defined for transactions whose transaction inputs reference at least
two different addresses since only those transactions can result in
the merging of address clusters. It assigns a value of one to the
transactions that merge two or more trivial address clusters since
these may be avoidable: the payer may need to combine several
transaction output values. However, any other instance of merging
could have been avoided by generating new key pairs. Even in the
presence of low address novelties, merge avoidance~\cite{hearn-13} can
produce high address cluster novelties.

We plot the address cluster novelties for each blockchain using an SMA
that includes the last \num{1}\% of transactions (red lines). We again
observe that the address cluster novelties in the Dogecoin and Clam
blockchains are much lower than that in the Bitcoin blockchain. We
note that even if one blockchain maintains or increases its address
and address cluster novelties on an individual basis, it can be
adversely impacted by a blockchain with lower novelties with which it
shares addresses.

\section{The Impact of a Single Blockchain}\label{sec:single-blockchain-impact}

Does the address clustering in the Clam blockchain provide us with
additional information that could improve the address clusterings in
the other blockchains? To answer this question, we define a graph
called the \textit{co-cluster graph} for a set of blockchains: Each
vertex represents an address cluster in a blockchain. Each edge
between two vertices represents the maximal set of addresses that are
shared between the two corresponding address clusters. If the two
address clusters do not share any addresses then there is no edge
between the vertices.

\begin{figure}
  \begin{center}
    \includegraphics[width=\linewidth]{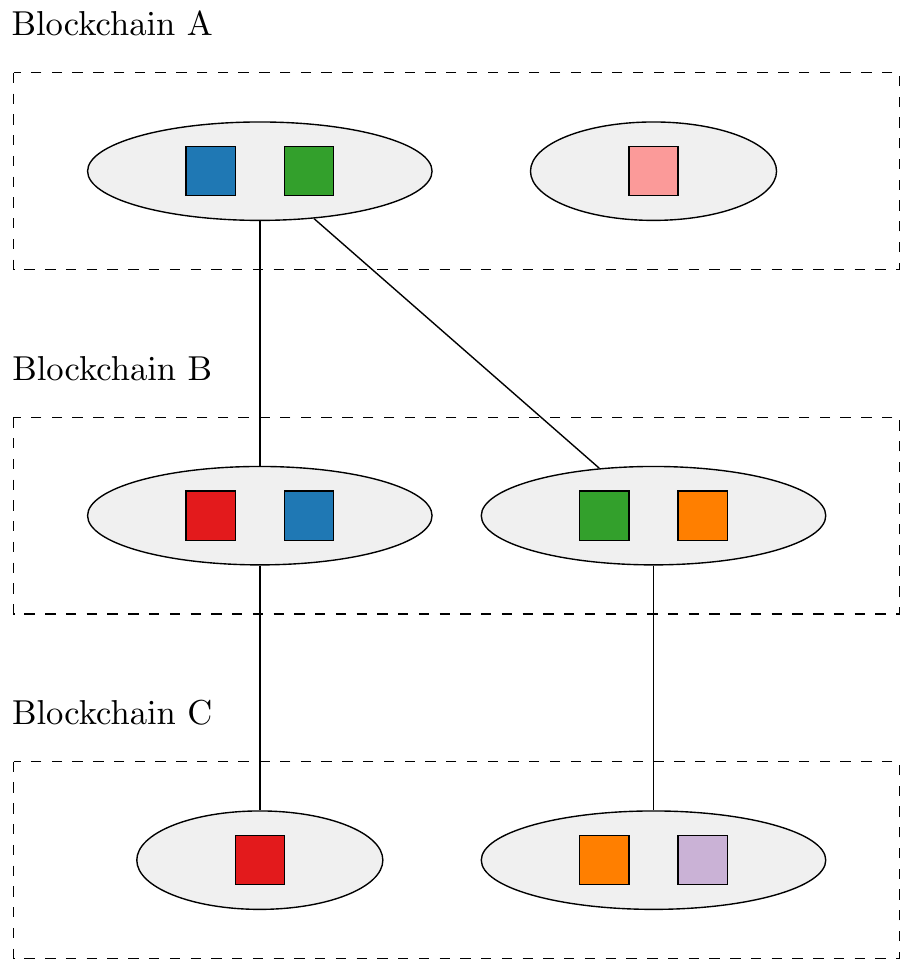}
  \end{center}
  \caption{The co-cluster graph for Blockchains~A, B and C\@. The six
    grey filled ellipses are the vertices and represent the address
    clusters in each blockchain. The four edges connect vertices whose
    address clusters share addresses (filled squares with the same
    colour) between the blockchains.}\label{fig:co-cluster-graph}
\end{figure}

\Cref{fig:co-cluster-graph} illustrates a co-cluster graph for three
blockchains. Blockchain~A (the topmost dashed rectangle) contains
three addresses (the blue, green and pink filled squares): two of the
addresses (blue and green) are in one address cluster and one address
(pink) is in another. The address clusters are represented by the grey
filled ellipses. Blockchain~B contains four addresses: two in one
address cluster and two in another. Two of the addresses (blue and
green) are shared between Blockchains~A and B\@. Blockchain~C contains
three addresses: one in its own address cluster and two in
another. Two of the addresses (red and orange) are shared between
Blockchains~B and C\@. No addresses are shared between Blockchains~A
and C\@.

The top-left vertex is incident with two vertices that correspond to
address clusters in the same blockchain. This is significant because
it shows that two addresses (blue and green) belong to the same
address cluster in Blockchain~A but different address clusters in
Blockchain~B\@. We can therefore merge the two address clusters in
Blockchain~B\@. In fact, it also indirectly shows that the two address
clusters in Blockchain~C can also be merged. We will return to this
point at the end of the section.

We can identify additional information that the Clam address
clustering provides over a counterpart by identifying the vertices
corresponding to address clusters in the Clam blockchain that are
incident with at least two vertices corresponding to address clusters
in the counterpart's blockchain. In other words, we wish to identify
address clusters in the Clam blockchain that contain addresses that
are shared with another blockchain but are found in two or more
address clusters in that blockchain.

We constructed the co-cluster graph for the Clam and Bitcoin
blockchains. We extracted the subgraph that is the maximal induced
subgraph on the set of vertices that represent Clam address clusters
with a degree of at least two, and their neighbours that represent
Bitcoin address clusters. This subgraph has \num{689} connected
components --- this number represents the number of `new and improved'
address clusters in the Bitcoin address clustering that are created by
the Clam address clustering. It represents a tiny fraction of ($\sim
0.0004$\%) of the original number of Bitcoin address clusters.

\begin{figure*}
  \includegraphics[width=\linewidth]{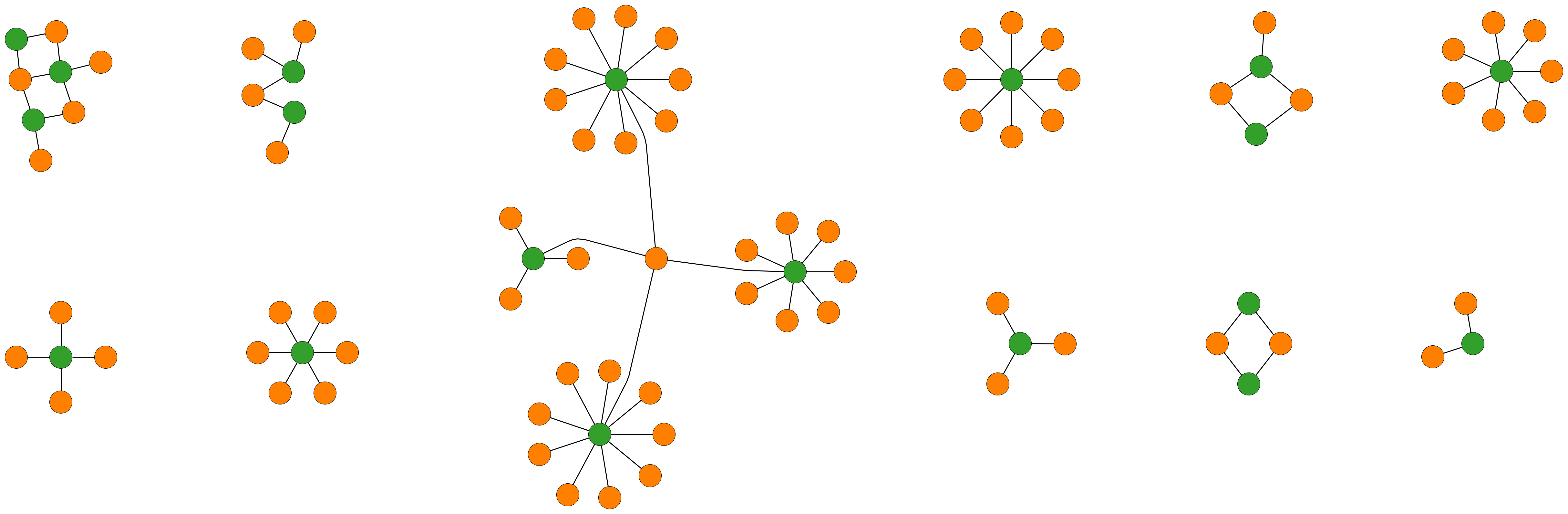}
  \caption{A visualisation of a portion of the subgraph of the
    co-cluster graph for the Clam and Bitcoin blockchains containing
    \num{689} connected components. The green and orange vertices
    represent address clusters on the Clam and Bitcoin blockchains,
    respectively.}\label{fig:bitcoin-clam-co-cluster-graph}
\end{figure*}

\Cref{fig:bitcoin-clam-co-cluster-graph} is a visualisation of a
portion of the extracted subgraph. The connected components can be
divided into stars and non-stars. The stars account for the vast
majority of the vertices. We hypothesise that stars represent an
entity transferring their Clams in one sweeping transaction. The
transaction merges the addresses on the Clam blockchain into a single
address cluster. However, the same addresses on the Bitcoin blockchain
belong to two or more address clusters. The non-stars are more
difficult to explain but they can be examined on an individual
basis. One possible explanation involves entities who purchased
disused private-keys in order to retrieve the corresponding Clam
balances.\footnote{https://bitcointalk.org/index.php?topic=2247688}

We repeated this analysis for the remaining two blockchains. The
extracted subgraph for the Litecoin blockchain has \num{179} connected
components. This represents a very small fraction ($\sim 0.0011$\%) of
the total number of Litecoin address clusters. In the case of
Litecoin, there is one large connected component that contains most of
the vertices (\num{2223} out of the \num{3255} vertices). The
extracted subgraph for the Dogecoin blockchain has \num{1181}
connected components.  This represents a small fraction ($\sim
0.0060$\%) of the total number of Dogecoin address clusters but it
is far higher than that observed in the Bitcoin or Litecoin
blockchains.

The analysis above measures the direct impact of the Clam address
clustering on the Bitcoin, Litecoin and Dogecoin blockchains. It shows
that the number of impacted address clusters is small but they can be
easily identified and examined on an individual basis. The analysis
can be repeated in the opposite direction, measuring the impact of the
Bitcoin, Litecoin and Dogecoin blockchains on the Clam blockchain, by
extracting the subgraphs in the opposite direction.

However, this method does not capture all interactions between a set
of blockchains. In \cref{fig:co-cluster-graph} the address cluster
represented by the top-left vertex indirectly impacts Blockchain~C's
address clustering via Blockchain~B\@. We can handle this by analysing
A's impact on B and then the impact of this new address clustering on
C\@. Alternatively, we can consider all three blockchains
simultaneously. We describe this method in the next section.

\section{The Impact of Multiple Blockchains}\label{sec:multi-blockchain-impact}

In \cref{sec:four-blockchains} we considered the impact of the Clam
address clustering on each blockchain in isolation. However, we can
combine blockchains to produce a single ordering of all of their
transactions. We can use the timestamps in the block headers to
produce an ordering that is approximately temporal. If we have access
to the times at which blocks and/or transactions were first broadcast,
we can improve the temporal accuracy of the ordering, but this is not
necessary for our purposes.

\begin{figure*}
  \centering
  \begin{subfigure}[c]{.45\linewidth}
    \includegraphics[width=\linewidth]{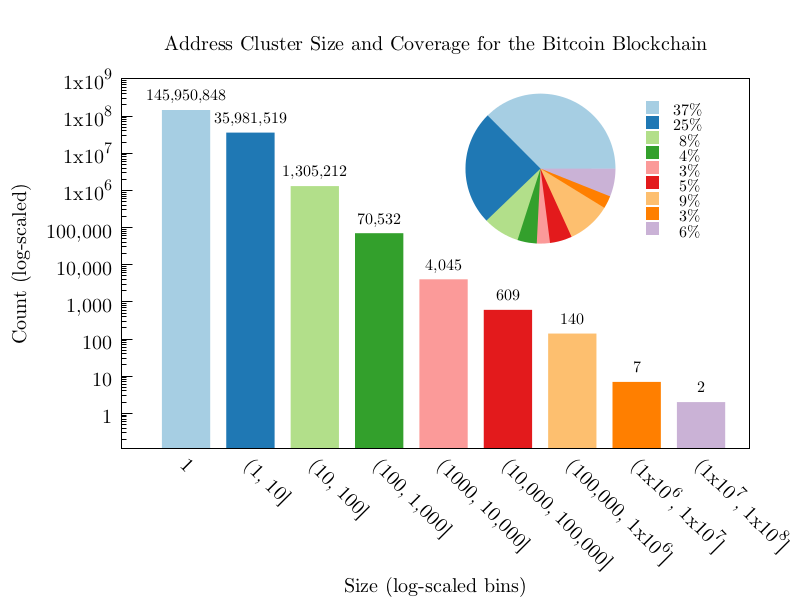}
    \caption{}\label{fig:bitcoin-cluster-analysis}
  \end{subfigure}
  \begin{subfigure}[c]{.45\linewidth}
    \includegraphics[width=\linewidth]{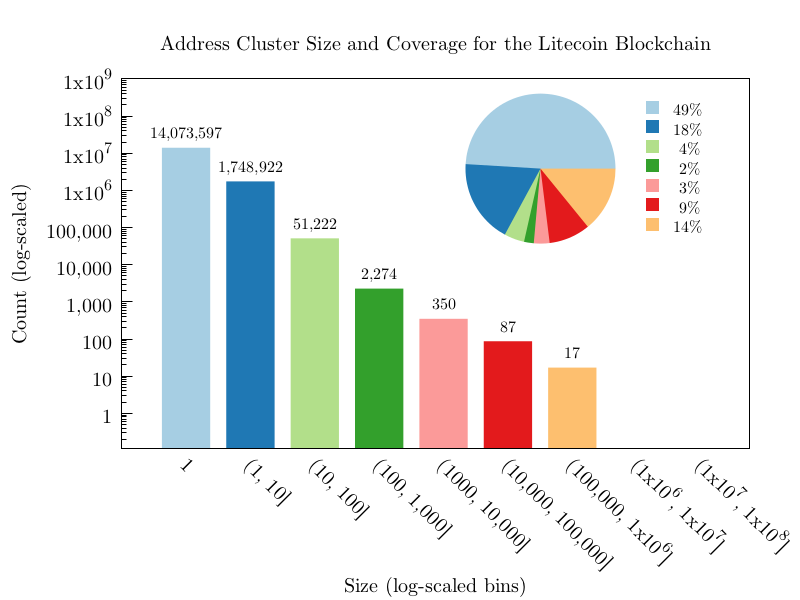}
    \caption{}\label{fig:litecoin-cluster-analysis}
  \end{subfigure}
  \begin{subfigure}[c]{.45\linewidth}
    \includegraphics[width=\linewidth]{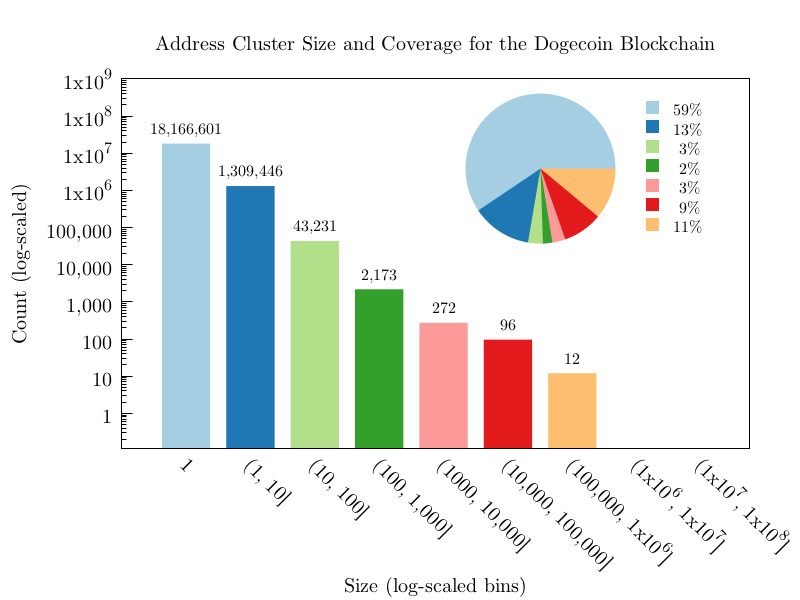}
    \caption{}\label{fig:dogecoin-cluster-analysis}
  \end{subfigure}
  \begin{subfigure}[c]{.45\linewidth}
    \includegraphics[width=\linewidth]{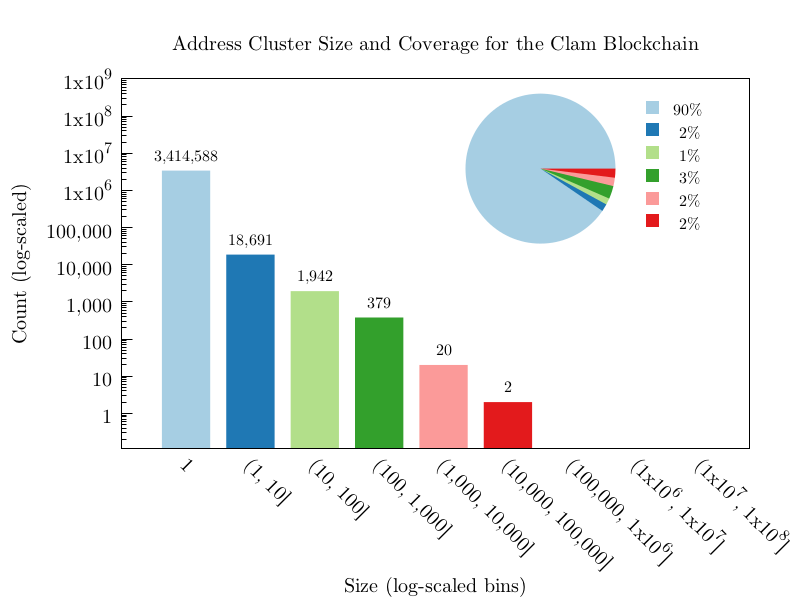}
    \caption{}\label{fig:clam-cluster-analysis}
  \end{subfigure}
  \begin{subfigure}[c]{.45\linewidth}
    \includegraphics[width=\linewidth]{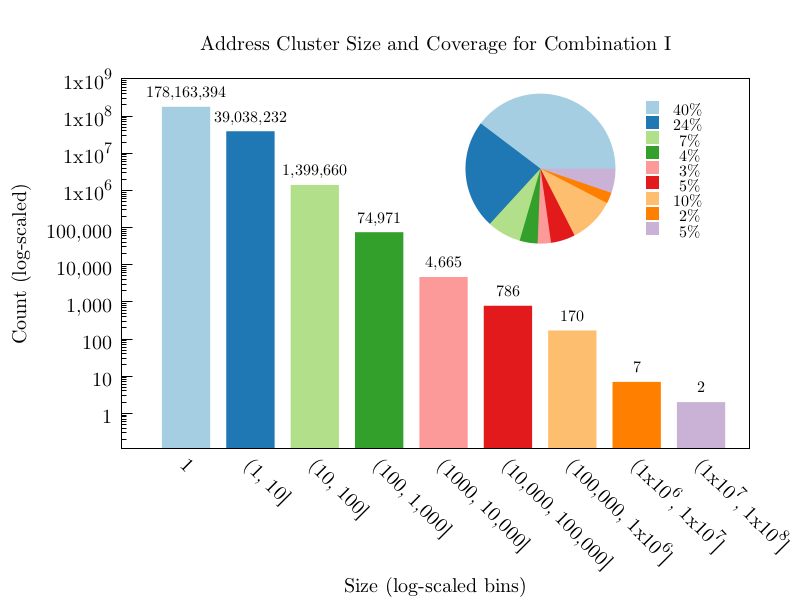}
    \caption{}\label{fig:bitlitedoge-cluster-analysis}
  \end{subfigure}
  \begin{subfigure}[c]{.45\linewidth}
    \includegraphics[width=\linewidth]{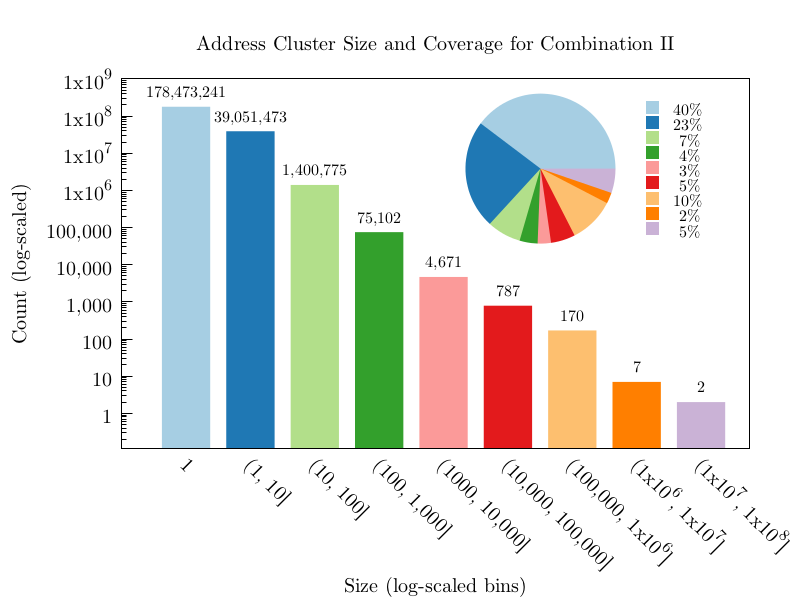}
    \caption{}\label{fig:bitlitedogeclam-cluster-analysis}
  \end{subfigure}
  \caption{Histograms showing the number of address clusters in each
    size range for the four blockchains and two combinations. The
    inset pie charts show the coverage, or total number of addresses
    in the address clusters, in each size range.}\label{fig:cluster-analysis}
\end{figure*}

We constructed two combinations: Combination~\rnum{1} included all
transactions in the Bitcoin, Litecoin and Dogecoin blockchains;
Combination~\rnum{2} included all transactions in the Bitcoin,
Litecoin, Dogecoin and Clam blockchains. We computed the sizes and
coverage of the address clusters (see
\cref{fig:bitlitedoge-cluster-analysis,fig:bitlitedogeclam-cluster-analysis})
and the novelties of the addresses and address clusters (see
\cref{fig:bitlitedoge-tx-analysis-annotated,fig:bitlitedogeclam-tx-analysis-annotated})
for the combinations as before. The dominant size of the Bitcoin
blockchain is evident. The difference between the two histograms is
marginal. However, the impact of the Clam airdrop on the address
novelties during the six month period following the creation of the
Clam genesis block is evident --- see the grey rectangles in the two
line charts. The airdrop decreases the address novelty in
Combination~\rnum{2} since the addresses were already observed in the
other three blockchains. We can also observe a decrease in the address
cluster novelty in Combination~\rnum{2} that occurs shortly after a
record price for Clam, as denominated in US dollars, was set in early
2018 --- see the grey circles in the two line charts. This may
indicate a renewed interest in the Clam airdrop and a resulting loss
in privacy.

\begin{figure}
  \begin{center}
    \includegraphics[width=\linewidth]{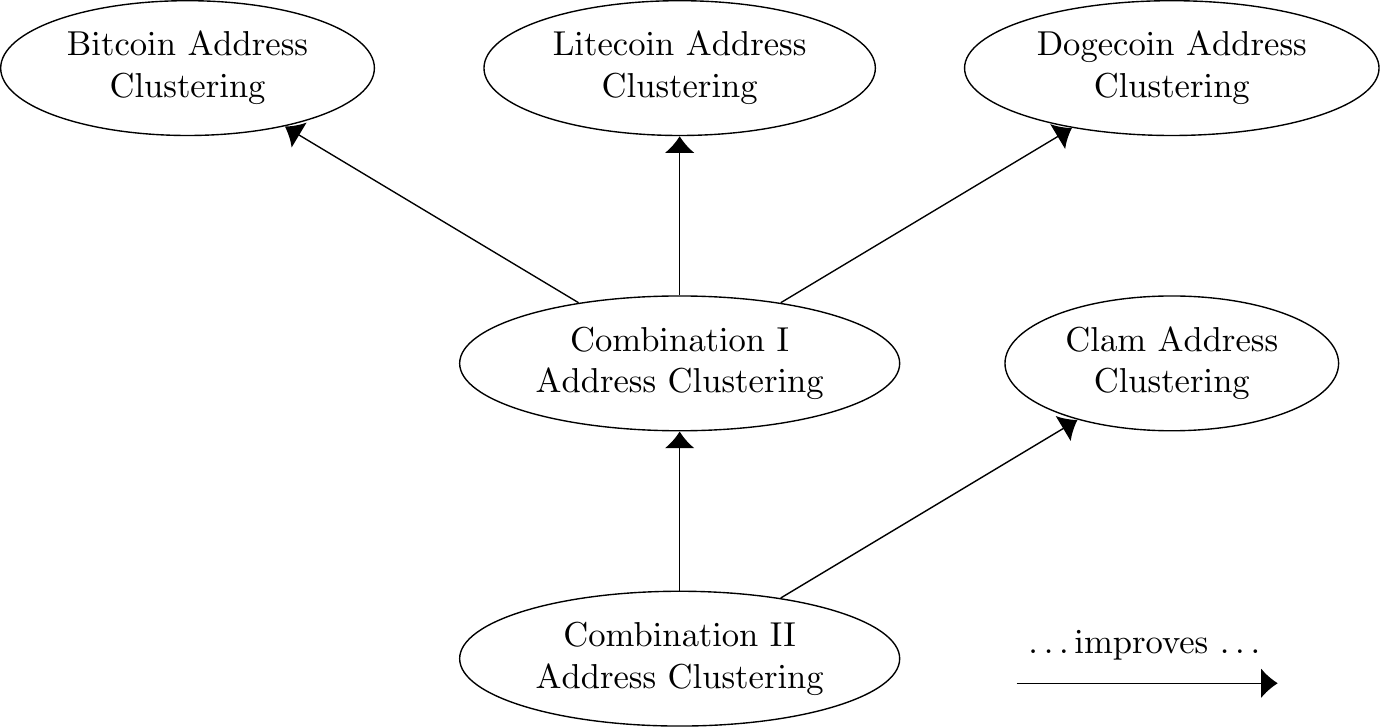}
  \end{center}
  \caption{A Hasse Diagram showing the improvement relation between
    the address clusterings for the individual blockchains and the
    combinations. The vertices represent the address clusterings and
    there is a directed edge from a source to a target if the address
    clustering corresponding to the source is an improvement of the
    address clustering corresponding to the
    target.}\label{fig:cluster-improvement}
\end{figure}

The address clusterings for the individual blockchains and the
combinations are related as follows. A partition $\Rho_1$ is an
\textit{improvement} of a partition $\Rho_2$ if and only if every
component of $\Rho_2$ is a subset of some component of $\Rho_1$ or is
disjoint with every component of $\Rho_1$. The address clustering for
Combination~\rnum{1} is an improvement of each of the address
clusterings for the Bitcoin, Litecoin and Dogecoin
blockchains. Similarly, the address clustering for
Combination~\rnum{2} is an improvement of both the address clustering
for Combination~\rnum{1} and the address clustering for the Clam
blockchain. It is also an improvement of the address clusterings for
the remaining blockchains since the relation is transitive. In fact,
the relation is a partial order and be visualised using a Hasse
Diagram (see \cref{fig:cluster-improvement}).

This relation provides a convenient method of measuring the impact of
the address clustering for the Clam blockchain. We can identify the
address clusters in Combination~\rnum{2} that are the union of two or
more address clusters in Combination~\rnum{1}. These are the
aggregated address clusters across the Bitcoin, Litecoin and Dogecoin
blockchains that the address clustering for the Clam blockchain has
improved or merged. We enumerated these address clusters: there are
\num{2161} of them. \Cref{sec:single-blockchain-impact} identified a
total of \num{2049} address clusters on the individual blockchains
that were directly impacted by the Clam address clustering. This new
number includes address clusters that are directly \textit{and
  indirectly} impacted by the Clam address clustering as described at
the end of the previous section.

\begin{figure}
  \begin{center}
    \includegraphics[width=\linewidth]{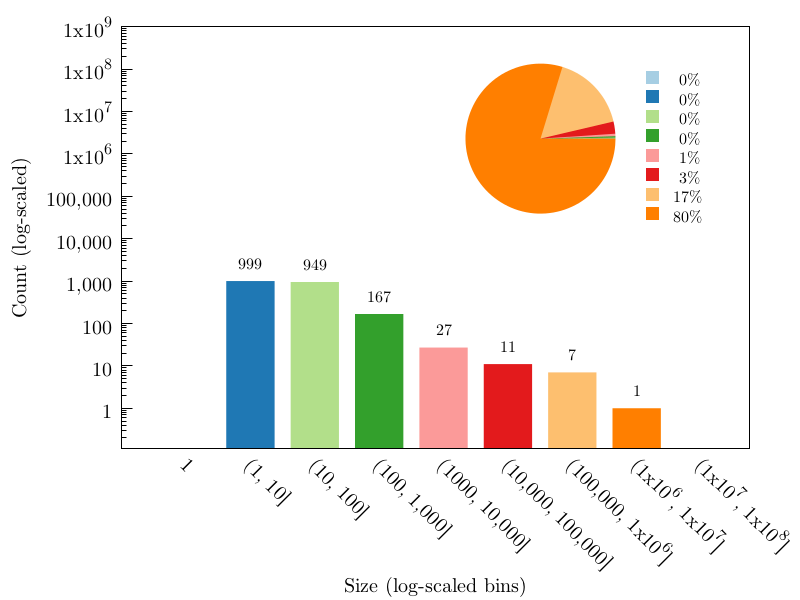}
  \end{center}
  \caption{A histogram showing the address cluster size and coverage
    for the address clusters for Combination~\rnum{2} that are the
    union of two or more address clusters for
    Combination~\rnum{1}.}\label{fig:combination-diff}
\end{figure}

\Cref{fig:combination-diff} shows the address cluster size and
coverage for the \num{2161} address clusters. Intuitively, this is the
difference between the histograms in
\cref{fig:bitlitedoge-cluster-analysis,fig:bitlitedogeclam-cluster-analysis}. The
Clam address clustering impacts both small and large address
clusters. Small address clusters may correspond to individuals making
their claims; large address clusters may correspond to centralised
services such as exchanges and mining pools making their claims. For a
more thorough analysis, one would need to use address tagging to
associate real-world identities with the address clusters. It is
important to note, that in both cases, the entities have disclosed
information about their address ownership on the Bitcoin, Litecoin and
Dogecoin blockchains, exclusively via their activity on the Clam
blockchain.

\section{Conclusion and Future Work}\label{sec:conclusion}

We have examined a privacy risk in using existing blockchains as a
basis for an airdrop's distribution. Specifically, we computed address
clusterings for the four blockchains involved in the Clam airdrop:
Bitcoin, Litecoin, Dogecoin and Clam. We showed that the sharing of
addresses between the blockchains leads to instances where an entity
discloses information about their address ownership on one blockchain,
exclusively via their activity on another. In the case of the Clam
airdrop, we identified $\sim 2000$ such instances.

The results can help blockchain analysts gather information regarding
the beneficiaries of airdrops. The opposing camp, those seeking to
hinder blockchain analysts, can also use the results to improve coin
control features that consider the implications of using addresses
that are shared across blockchains.

Our future work centres around applying this analysis to other
airdrops. These include \textit{holder airdrops} similar to the
airdrop discussed in this paper, e.g.\ Bitcoin Private, BitCore and
United Bitcoin, and \textit{forked airdrops}, e.g.\, Bitcoin Cash,
Bitcoin Gold and Bitcoin Diamond. A fork creates an airdrop since the
entire set of addresses at the point of the fork are shared between
the resulting blockchains. A user's activity on one side of the fork
can expose information about their address ownership on the other. We
would also like to generalise this approach to study airdrops
involving Ethereum-based tokens.

\bibliographystyle{plain}

\bibliography{bib/main}

\end{document}